\documentclass[modern,12pt]{aastex62}
\usepackage{newtxtext,newtxmath} 
\usepackage[USenglish]{babel}
\usepackage[utf8]{inputenc}
\usepackage[T1]{fontenc}
\usepackage{comment}
\usepackage{enumitem}
\usepackage{glossaries}
\usepackage{todonotes}
\usepackage{collect}
\usepackage{wrapfig}
\usepackage{titlesec}

\setlist{noitemsep}
\graphicspath{{./}{./figures/}{./tabs}}
\newcommand{\Contact}[1]{}

\providecommand{\secref}[1]{\hyperref[#1]{Section~\ref{#1}}}

\providecommand{\recref}[1]{\hyperref[#1]{REC-\ref{#1}}}
\newcounter{reccount} 
\newenvironment{recenv}[1]
    { 
    \refstepcounter{reccount}  
    #1
    }

\newcommand{\nrec}[6]{ 
	\begin{recenv}
	\vspace{5pt}
	\noindent \textbf{REC-\thereccount:~#5}. \newline
	\textbf{Area:} #1.  \textbf{Audience:} #2. \textbf {Time Frame:} #3
	\newline \label{rec:#4}\noindent \textit{#6}
	\vspace{5pt}
	\end{recenv}
\typeout{REC-\thereccount: #1: #2: #3: #4: #5 :END}
}

\makeatletter
\renewcommand{\section}{ \@startsection
{section}
{1}
{0mm}
{-1.5ex \@plus -1ex \@minus -.2ex}
{0.3ex \@plus.2ex}
{\normalfont\bfseries}}
\makeatother

\begin{document}
\title{\large Algorithms and Statistical Models for \\ Scientific Discovery in the Petabyte Era \\
\footnotesize
Astro2020 APC White Paper: \hspace*{3pt} Thematic Areas -\hspace*{2pt} $\boxtimes$ Activity \hspace*{15pt} $\square$ Project \hspace*{15pt} $\boxtimes$ State of the Profession 
}




\small
\author{Brian~Nord}
\affiliation{Fermi National Accelerator Laboratory}
\affiliation{Kavli Institute of Cosmological Physics, University of Chicago}
\affiliation{Department of Astronomy and Astrophysics, University of Chicago}
\author{Andrew~J.~Connolly}
\affiliation{DIRAC Institute, Department of Astronomy, University of Washington}
\author{Jamie~Kinney}
\affiliation{Google, Inc.}
\author{Jeremy~Kubica}
\affiliation{Google, Inc.}
\author{Gautaum~Narayan}
\affiliation{Space Telescope Science Institute}
\author{Joshua~E.~G.~Peek}
\affiliation{Space Telescope Science Institute}
\affiliation{Department of Physics \& Astronomy, The Johns Hopkins University}
\author{Chad~Schafer}
\affiliation{Carnegie Mellon University}
\author{Erik~J.~Tollerud}
\affiliation{Space Telescope Science Institute}


\author{Camille~Avestruz}
\affiliation{Enrico Fermi Institute, The University of Chicago}
\affiliation{Kavli Institute of Cosmological Physics, University of Chicago}

\author{G.~Jogesh~Babu}
\affiliation{Center for Astrostatistics, Penn State University}

\author{Simon~Birrer}
\affiliation{Department of Physics \& Astronomy, University of California, Los Angeles}

\author{Douglas~Burke}
\affiliation{Center for Astrophysics, Harvard \& Smithsonian}

\author{Jo\~ao~Caldeira}
\affiliation{Fermi National Accelerator Laboratory}

\author{Douglas~A.~Caldwell}
\affiliation{SETI Institute}

\author{Joleen~K.~Carlberg}
\affiliation{Space Telescope Science Institute}

\author{Yen-Chi~Chen}
\affiliation{Department of Statistics, University of Washington}

\author{Chuanfei~Dong}
\affiliation{Princeton University}

\author{Eric~D.~Feigelson}
\affiliation{Department of Astronomy \& Astrophysics, Penn State University}
\affiliation{Center for Astrostatistics, Penn State University}

\author{V.~Zach~Golkhou}
\affiliation{DIRAC Institute, Department of Astronomy, University of Washington}

\author{Vinay~Kashyap}
\affiliation{Center for Astrophysics, Harvard \& Smithsonian}

\author{T.~S.~Li}
\affiliation{Fermi National Accelerator Laboratory}

\author{Thomas Loredo}
\affiliation{Cornell University}

\author{Luisa~Lucie-Smith}
\affiliation{Department of Physics \& Astronomy, University College London}

\author{Kaisey~S.~Mandel}
\affiliation{University of Cambridge}

\author{J.~R.~Mart\'{i}nez-Galarza}
\affiliation{Center for Astrophysics, Harvard \& Smithsonian}

\author{Adam~A.~Miller}
\affiliation{CIERA, Northwestern University}
\affiliation{The Adler Planetarium}

\author{Priyamvada~Natarajan}
\affiliation{Department of Astronomy, Yale University}

\author{Michelle~Ntampaka}
\affiliation{Harvard Data Science Initiative, Harvard University}
\affiliation{Center for Astrophysics, Harvard \& Smithsonian}

\author{Andy~Ptak}
\affiliation{NASA/Goddard Space Flight Center}

\author{David~Rapetti}
\affiliation{CASA, University of Colorado, Boulder}
\affiliation{Department of Astrophysical \& Planetary Science, University of Colorado, Boulder}
\affiliation{NASA Ames Research Center}

\author{Lior~Shamir}
\affiliation{Department of Computer Science, Kansas State University}

\author{Aneta~Siemiginowska}
\affiliation{Center for Astrophysics, Harvard \& Smithsonian}

\author{Brigitta~M.~Sip\H{o}cz}
\affiliation{DIRAC Institute, Department of Astronomy, University of Washington}

\author{Arfon~M.~Smith}
\affiliation{Space Telescope Science Institute}

\author{Nhan~Tran}
\affiliation{Fermi National Accelerator Laboratory}

\author{Ricardo~Vilalta}
\affiliation{Department of Computer Science, University of Houston}

\author{Lucianne~M.~Walkowicz}
\affiliation{The Adler Planetarium}
\affiliation{LSSTC Data Science Fellowship Program}

\author{John~ZuHone}
\affiliation{Smithsonian Astrophysical Observatory}

\correspondingauthor{Brian~Nord}
\email{nord@fnal.gov}


\maketitle







\newpage
\normalsize
\begin{center}
    \large 
    Abstract
    \normalsize
\end{center}
\vspace{-5pt}
The field of astronomy has arrived at a turning point in terms of size and complexity of both datasets and scientific collaboration. 
Commensurately, algorithms and statistical models have begun to adapt --- e.g., via the onset of artificial intelligence --- which itself presents new challenges and opportunities for growth.
This white paper aims to offer guidance and ideas for how we can evolve our technical and collaborative frameworks to promote efficient algorithmic development and take advantage of opportunities for scientific discovery in the petabyte era.
We discuss challenges for discovery in large and complex data sets; challenges and requirements for the next stage of development of statistical methodologies and algorithmic tool sets; how we might change our paradigms of collaboration and education; and the ethical implications of scientists' contributions to widely applicable algorithms and computational modeling. 
We start with six distinct recommendations that are supported by the commentary following them.
This white paper is related to a larger corpus of effort that has taken place within and around the Petabytes to Science Workshops \citep[\url{https://petabytestoscience.github.io/};][]{2019arXiv190505116B}. 

\section{Strategic Plan: Recommendations}
    \nrec{Analysis}{Agency, Astronomer}{Medium}{algmodels}{
    Create funding models, opportunities, and programs to support the development of advanced algorithms and statistical methods specifically targeted to the astronomy domain} {The increasingly large and complex datasets resulting from a new generation of telescopes, satellites, and experiments require the development of sophisticated and robust algorithms and methodologies.
    These techniques must have statistically rigorous underpinnings, as well as being adaptable to changes in computer architectures. It is also critical to fund the long-term maintenance, curation, and distribution of these algorithms.}
    \nrec{Analysis}{Technologist, Astronomer}{Long}{discengine}{
    Build automated discovery engines}
    {New hypotheses are difficult to generate in an era of large and complex datasets.
    Frameworks that can detect outliers or new patterns (and reliably minimize false-positive) within our data could address many of the needs of current and planned science experiments.
    Funding and developing these engines as a community would lead to broad access to the tools needed for scientific exploration.}
    \nrec{Analysis}{Agency, Manager, Astronomer}{Long}{algconnect}{
    Promote interdisciplinary collaboration between fields, and industry}
    {Expertise across multiple domains is required to tailor algorithmic solutions to astronomical challenges.
    The astronomical community should more heavily and directly engage researchers from industry and non-astronomy fields in the development and optimization of algorithms and statistical methods.
    Agencies and academic departments should develop  programs to specifically connect astronomers to these experts through sabbatical programs, centers, fellowships, and workshops for long-term cross-domain embedding of experts.
    }
    \nrec{Analysis}{Agency, Astronomer}{Medium}{algeducation}{
    Develop an open educational curriculum and principles for workforce training in both algorithms and statistics}
    {The speed of model and algorithm evolution requires regular training and education for scientists and for those seeking to enter science.
    Developing and maintaining open curricula and materials would enable the teaching of algorithms and methodologies throughout the astronomical community.}
    \nrec{Analysis}{Astronomer, Agency}{Short}{algpub}{
    Encourage, support, and require open publication and distribution of algorithms}
    {The rapid adoption of advanced methodologies and the promotion of reproducible science would be significantly enhanced if we mandated the open publication and distribution of algorithms alongside papers.}
    \nrec{Ethics}{Astronomer, Agency}{Long}{algethics}{
     Engage and facilitate community-wide conversations and working groups to discuss the societal applications and ethical implications of algorithms and software to which astronomers contribute}
    {Advanced algorithms have broad applicability outside strictly STEM contexts. Astronomers are poised to become more involved in the development of these algorithms and thus have a role in the discussion regarding how they are deployed in society. The astronomy community should take action to develop a strategic plan for its role in addressing the societal implications of these algorithms.}

\section{Introduction}
The paradigms for data analysis, collaboration, and training have simultaneously reached a watershed moment in the context of algorithms and statistical methods.
The onset of large datasets as a scientific norm accentuates this shift, bringing both technical opportunities and challenges.
As we address these challenges with new algorithmic approaches,
the incorporation of rigorous statistical paradigms must keep apace.
We as a community, however,  lack many of the tools to even contend with, much less  take full advantage of, increasingly complex datasets for discovery.

These paradigm shifts  bring organizational challenges that highlight issues with the cultural norms for education and collaboration. 
Discovery often occurs at the intersections of, or in the interstices between domains, and therefore
multi-disciplinary collaboration has  become a key component of research.
We need to improve these methods of collaboration to take advantage of the rapid emergence of technologies and to increase the permeability of the barrier between different areas of science, and between academia and industry.
Moreover, innovation in methods of education and training 
lag behind the development of the techniques themselves, leading to a growing disparity in the distribution of knowledge.
Similarly,  software development strategies must keep apace with these developments, both to ensure results are robust and that education and training are  equitably distributed.

We have an opportunity to 
leverage our community's energy and inspiration to initiate change in how we drive algorithmic discovery in the petabyte era.
Below, we discuss the key challenge areas 
and provide possible directions for what we can do.

\section{Discovery in the Petabyte Era}
Currently,
hypothesis generation in astronomy has two pathways.
One is theoretical, wherein predictions from theory provide hypotheses that can be tested with observations.
The other is observational, wherein surprising objects and trends are discovered in data and later explored.
As theory comes to depend on large-scale simulations and observational datasets grow into and beyond the petabyte scale, both of these pathways are coming under threat.
Classical modes of data exploration are becoming prohibitively slow
(e.g. finding objects and correlations by plotting parameters within datasets).

 A key example of the challenge lies in explorations of high-dimensional datasets.
Long ago, the discovery that stars fill an approximately 1D space in magnitude-color space led to a physical model of stellar structure.
This is a low-dimensional, non-linear representation of higher-dimensional data.
Indeed, seemingly smooth structures in astronomical data can have surprising substructure \citep[e.g., the Jao/Gaia Gap][]{Jao_2018}).
1D gaps in famous 2D spaces are visually discoverable.
However, we lack comparable methods to find 2D gaps in 3D spaces, let alone structures in the extremely high-dimensional data that modern surveys create.
Recently, \cite{2018AJ....156..219S} found 17 pure blackbody stars \emph{by eye} amongst the 798,593 spectra in SDSS, nearly two decades after they were acquired.
This result shows both how interesting outliers can be, and how by-eye methods are impractical
at the petabyte scale.
With trillions of rows available in upcoming surveys, we'll have the ability to find low-dimensional substructure in high-dimensional that has potential to yield new physical insight --- but only if we have the tools to do so.

As an example of such a tool, purpose-built Machine Learning (ML) algorithms coupled with deep sub-domain knowledge can successfully expose hitherto unknown objects that can significantly advance our understanding of our universe \citep[e.g.][]{2017MNRAS.465.4530B}.
Unfortunately, any successful exploration requires a) deep algorithmic and implementation knowledge b) deep physical and observational domain knowledge and c) luck.
Deep algorithmic knowledge is necessary as off-the-shelf algorithms usually need significant adaptation to work with heteroscedastic and censored astronomical data.
Deep observational domain knowledge is needed as outlier objects are often artifacts and surprising trends may be imprints of the data collection method.
Deep physical domain knowledge is needed to make sense of the result, and understand its place in the cosmos.
Finally, not all searches will return results; a modicum of luck is needed.
This trifecta of algorithmic knowledge, domain knowledge, and luck is rare.



We argue that the path forward is through the construction of intuitive, trustworthy, robust, and deployable algorithms that are intentionally designed for the exploration of large, high-dimensional datasets in astronomy.
When we consider the current landscape of astronomical research and the upcoming generation of sky surveys, we identify areas where algorithmic and statistical investment are needed, such as a) online/real-time alerts and anomaly detection in large sky surveys; b) statistical learning models that are physically interpretable; c) physics-aware machine learning algorithms.

Very few researchers have both all the needed skills and the bravery/foolhardiness to seek out risky avenues of research like these.
We therefore propose that funding agencies fund opportunities for the creation and maintenance of ``discovery engines'' --- tools that allow astronomers without deep algorithmic knowledge to explore the edges of data spaces to hunt for outliers and new trends.
These engines should be hosted near the data, but should be initiated by the astronomical and methods-development communities.

The development of new statistics and algorithms can be accomplished through a variety of methods, including: on-boarding dedicated algorithmic/data-intensive science experts onto astronomy teams, facilitating partnerships (with industry or other academic fields), and building internal expertise within the community through education and training.
Regardless of the mechanism, it is important that the development of new statistical and algorithmic techniques is considered a core part of astronomical missions.

\section{The state of statistics: methodologies for astrophysics}

Statistical methods and principles are the backbone upon which successful estimation, discovery, and classification tasks are constructed.
The tools commonly associated with Machine Learning (e.g. deep learning) are typically efficient, ``ready-to-use" algorithms (albeit with ample tuning parameters).
On the other hand, statistical approaches employ a set of data analysis principles.
For example, Bayesian and frequentist inference are two competing philosophical approaches to parameter estimation, but neither prescribes the use of a particular algorithm.
Instead, the value (and perhaps the curse) of the statistical approach is that methodological choices can be tailored to the nuances and complexities of the problem at hand.
Hence, when considering the statistical tools that are crucial for astronomy in the coming decade, one must think of the recurring challenges that are faced by data analysis tasks in this field.

Further, as the sizes of astronomical survey datasets grow, it is not sufficient to merely “scale up” previously-utilized statistical analysis methods.
More precisely, modern data are not only greater in volume, but are richer in type and resolution.
As the size and richness of datasets increase, new scientific opportunities arise for modeling known phenomena in greater detail, and for discovering new (often rare) phenomena.
But these larger datasets present challenges that go beyond greater computational demands: they are often of a different character due to the growing richness, which necessitates new analysis methods and therefore different statistical approaches.
Hence, as the complexity of astronomical data analysis challenges grow, it is imperative that there be increasing involvement from experts in the application of statistical approaches.

To ground these ideas, we consider examples of technical and organizational challenges that, if advanced over the next decade, would significantly benefit  astronomy.\\ \\
\noindent {\it Technical Challenges}
\vspace{-6pt}
\begin{enumerate}
\item {\it Methods for the analysis for noisy, irregularly-spaced time series.}
Future time domain surveys, like LSST,  will generate a massive number of light curves (time series) with irregular observational patterns and in multiple bands.
\item {\it Likelihood-free approaches to inference.}
 Likelihood-free approaches, such as approximate Bayesian computation, hold promise for astronomical applications. Recent work has developed tools and proofs of concept for astronomy datasets to demonstrate that this computationally-intensive approach is feasible \citep[e.g.,][]{2018MNRAS.477.2874A}.
\item {\it Efficient methods of posterior approximation.}
Even in cases where a likelihood function is available, constructing the Bayesian posterior is challenging in complex cosmological parameter estimation problems, because future inference problems will push the computational boundaries of current MCMC samplers.
\item {\it Emulators for complex simulation models.}
It is increasingly the case that simulations provide the best understanding of the relationship between  parameters of interest and the observable data. 
Unfortunately, these simulation models are often of sufficient complexity that a limited number of simulation runs can be performed; the output for additional input parameter values must be approximated using emulators that interpolate these available runs. 
\item {\it Accurate quantification of uncertainty.}
Complex inference problems in astronomy are often, out of necessity, divided into a sequence of component steps.
This divide-and-conquer approach requires careful consideration of the propagation of error through the steps.
How does one quantify errors in such a way that these uncertainties are accurately accounted for in the downstream analyses?
\end{enumerate}

\noindent {\it Organizational Challenges}
\vspace{-6pt}
\begin{enumerate}
\item {\it Accessible publishing of methods.}
Advances in statistical theory and methods abound in the literature of that field, but it is often presented in a highly formalized mathematical manner, which obscures the aspects of most importance to potential users.
This creates a barrier to the appropriate use of these methods in astronomy.
This will require appropriate professional recognition for this work, including encouraging the publication of methodology papers in astronomical journals by data scientists.

\item {\it Avoiding the ``algorithm trap."}
Astronomical inference problems are of sufficient complexity that full use of the data requires analysis methods to be adapted and tailored to a specific problem.
For this reason, statisticians prefer to not think of an analysis as the application of a ready-made algorithm.
Astronomers, however, are generally more interested in the result of the analysis, so are attracted to an algorithm they can apply.
This difference in perspective only increases the need to have data scientists deeply involved in the collaborative process to help find the best matches between scientific problem and algorithm/statistical model.
\item {\it Reducing barriers for statisticians.}
From the other side, data scientists face challenges in applying analysis techniques to astronomical data.
This is partially driven by the fact deeply understanding the science is often crucial to building methods tailored to the problem. 
More effort is needed to reduce these barriers.
\end{enumerate}

\section{The state of algorithms: developments in the last decade}

A key milestone for enabling and accelerating science in this past decade 
has been the development of 
general-purpose algorithms that can be applied to a variety of problems. These algorithms, irrespective of what programming language the implementation is in, have made astrophysical research more repeatable and reproducible, and less dependent on human tuning.


As with statistics, the distinction between algorithms, and the software implementation of algorithms is blurry within the community. In many situations, we now use algorithms without any knowledge of how they work.
For example, we can now expect to sort tables with millions of rows on multiple keys, without knowing the details of sorting algorithms, precisely because these details have been abstracted away.
We note that the many widely used algorithms, such as affine-invariant Markov Chain Monte Carlo techniques are widely used precisely because the algorithm is implemented as a convenient software package.
Community-developed software packages such as \texttt{scikit-learn}, \texttt{astropy}, and the IDL Astronomy Library have increased the community's exposure to various algorithms, and the documentation of these packages has in many cases supplanted implementation-oriented resources such as Numerical Recipes.

At the same time in the broader world, a class of algorithms is being used to execute tasks for which an explicit statistical forward model is too complex to develop, and correlations within the data itself is used to generate actionable predictions.
These AI techniques include machine learning models, which have been used to replace humans for tasks as varied as identifying artifacts in difference images, to categorizing proposals for time allocation committees.
Deep learning methods, in particular, are increasingly viewed as a solution to specific petabyte-scale problems, as they have been successfully deployed in the commercial sector on these scales.
We anticipate increasing adoption of these algorithms, as user-friendly implementations such as \texttt{pyTorch} and \texttt{Keras} become more well known, and data volumes grow.
It is also likely that the algorithms that are used to train these machine learning methods, including techniques like stochastic gradient descent, will find more use within the astronomical community.

Machine learning algorithms are necessary but not sufficient to continue the progress in astrophysical research that is driven by algorithms.
In particular, machine learning methods are often not interpretable, and while their output can be used effectively, those outputs are not true probabilities. The scientific method fundamentally involves the generation of a testable hypothesis that can be evaluated given data, and is therefore inherently statistical. As data volumes grow, the dimensionality of models grows, and there is increasing recognition that the model structure is hierarchical or multi-level. While we see increasing adoption of hierarchical models for Bayesian inference, there remains much to do to increase awareness of algorithms to effectively evaluate these models, including probabilistic programming - algorithms that are used to build and evaluate statistical models in a programmatic manner.

\vspace{3pt}

\noindent {\it Technical challenges}
\vspace{-6pt}
\begin{enumerate}
\item Both algorithms and models need to be trustworthy and interpretable. 
It's easy to throw a dataset into a neural net or ensemble classifier and overfit.
\item Many algorithms, especially in the machine learning space, require labeled data that may not be available at sufficient volumes, or at all.
\item The reproducibility of results derived from algorithms needs to be improved. 
This is especially important with machine learning models where black-box optimization is often used because it is an easy-to-provide feature. 
\item Scalability of newly-developed algorithms. With the data volumes of the petabyte era, efficiency in all parts of the stack is necessary.  
\item Significant work is still needed in adapting and improving the current space of existing algorithms: optimizing traditional astronomy algorithms, adapting them for a cloud setting, or even making small accuracy improvements.
\end{enumerate}

\noindent {\it Organizational challenges}
\vspace{-6pt}
\begin{enumerate}
  \item It is difficult to get the necessary expertise onto all missions that will need it both in terms of developing the expertise internally (due to the fast pace of change in the space) and hiring in experts.
  \item There is currently no established marketplace/mechanism for matching difficult problems in the astronomy domain to relevant experts outside an astronomer's network.  
  \item There is a missing component in the conduit of moving new algorithms developed in academia into robust, usable, finished products. 
  \item We need standardized processes for publishing algorithms and machine learning models such that the results obtained with these algorithms/models are: broadly accessible, discoverable, fully reproducible (including archiving the model parameters), and easily comparable with other algorithms in the problem space.
  \item We need to define and fund a process for continually modernizing/upgrading algorithms as the broader environment changes (new languages, new libraries, new computational architectures, shift to cloud computing, etc). 
\end{enumerate}

\section{Emerging trends in industry and other fields}
\label{sec:algos:industrytrends}

Over the past two decades, the wider industry has also seen a shift in development approaches and computational techniques that can be adopted by the astronomical community.
Open source software has become a new normal with communities sharing their investment in software development.
When considered along with the industry's shift toward cloud computing and software-as-a-service, astronomy can benefit from the new scale and availability of off-the-shelf solutions for computation and storage.
Astronomers no longer need to focus significant portions of time on the low-level technical details in running dedicated banks of computers to support each survey.

This service model is being extended beyond software deployments and starting to push into algorithms as a service.
Cloud machine learning services provide a portfolio of general algorithms.
Instead of worrying about the specifics of the algorithm development, users focus only on model specification.
This requires a shift in how we think about new algorithm development.
Instead of focusing on the details, such as implementation, optimization, and numerical accuracy, the practitioner focuses primarily on the high level model specification.
Due to a series of recent successes, a significant focus within hosted machine learning services has been on deep neural networks (DNNs).
NNs have shown remarkable success across a variety of tasks.
Further new developments such as convolutional neural networks and recurrent neural networks have extended the power of this technique.

Another area of focus within the field of machine learning is blackbox optimization (e.g. AutoML).
AutoML systems aim to abstract away not just the algorithm’s implementation details, but also the need to manually tune model parameters.
For example, recent work in neural architecture search allows the AutoML system to handle  development decisions such as choosing the structure of the network (number and width of layers) as well as the learning parameters.
While this automation greatly simplifies the problem of constructing accurate models, it does move the practitioner one step further from understanding the full details of the model.

There is an opportunity for astronomy to both benefit from and help drive new advances in the emerging industries.
As noted above, astronomy can benefit from the shift from individually developed and maintained systems to hosted platforms that allow more effort to be spent on the data analysis itself.
Moreover, the shape and size of science data serve as a driver for the development of new algorithms and approaches.
We expect many of the upcoming advancements to be driven by real-world problems—machine learning will rise to the challenge of solving new, open problems.
The recommendations in this paper aim to ensure some of these problems and solutions are in the astornomy domain.


\section{Enhancing Interdisciplinary Programs and Collaborations}

The past decade has been a period of rapid change in the the multi-dimensional landscape of algorithms, computing, and statistics. We have seen the rise of new ``standard'' programming languages and libraries (e.g. Python, \texttt{astropy}, \texttt{scikit-learn}).
There has been a proliferation of new algorithmic and statistical techniques --- from improvements in image processing and compression to the rise of deep neural networks as a powerful tool from machine learning.
We have seen the rise of new computational modalities, such as cloud computing and software-as-a-service.
New distributed computing frameworks such as Dask and Spark are emerging to process and analyze large and complex datasets.
Even the basic mechanics of computation is undergoing a shift with the availability of specialized hardware such as GPUs and TPUs, requiring a new domain of knowledge to efficiently deploy solutions.
There is no reason to expect the pace of innovation to drop  anytime soon.

This rapid pace of advancement means that it is no longer possible for a single astronomer or even a small team of astronomers to build the necessary depth of expertise in all of these areas.
As new technologies spring up quickly,  the astronomical community will need to balance the cost of learning the new technologies with the benefits they provide.
It is not reasonable to expect every astronomer to keep up with all of the advances.
In response to this, a number of new ad hoc collaborations or collectives have sprung up to bring together astrophysicists and deep learning experts, such as the Deep Skies Lab (\url{deepskieslab.ai}) and Dark Machines (\url{darkmachines.org}).

In cases where collaborations exist today, there can be a variety of complicating challenges.
There is currently no established marketplace for matching difficult problems in the astronomy domain to relevant experts outside an astronomer's network (see also \S\ref{sec:algos:industrytrends}).
The resulting in-depth collaborations have start up overhead as the external experts learn enough about the problem domain to be helpful. 
Short-term engagements can suffer from a lack of depth or insufficiently productionized solutions.
Even in longer term engagements, there can be misalignment between the parties due to the different incentives.
For example, statisticians and computer scientists in academia are primarily recognized for only the novel contributions to their own fields.
Papers that apply existing methodologies to new problems are not considered significant contributions to their fields.
Similarly, members of the astronomy community are not fully recognized for their algorithmic contributions.

The form and depth of the engagement will naturally be project-dependent.
Experimental and privately-funded interdisciplinary centers e.g.\ the Moore-Sloan Data Science Environments at Berkeley, NYU and the University of Washington, or the Simon's Flatiron Institute have demonstrated how expertise in data science can advance a broad range of scientific fields.
Access to the resources at these centers is, however, limited to researchers at these privileged institutions.
The challenge we face is how to scale these approaches to benefit our community as a whole.

\section{Education and training}

Training a workforce that can address the algorithmic and statistical challenges described in this Paper will require a significant change in how we educate and train everyone in our field, from undergraduate students to PI's.
The traditional curricula of physics and astronomy departments do not map easily to the skills and methodologies that are required for complex and/or data intensive datasets.
This is a rapidly changing field, and will remain so for at least a decade.
However, a strong foundation in Bayesian statistics, data structures, sampling methodologies, and software design principles would enable professional astronomers to take advantage to big data in the next decade.
Bridging this gap between the skills we provide our workforce today and the ones they might need to succeed in the next decade should be a priority for the field.
In the previous decade there was substantial progress in creating material to support the teaching of statistics and machine learning in astronomy.
This includes the publication of introductory textbooks \citep{astroMLText,R:Kohl:2015en}, the creation of common software tools and environments, the development of tutorials, and a growing focus on software documentation \citep{astropy_tutorials}.
The emergence of Jupyter \citep{jupyter} as a platform for publishing interactive tutorials and Github and Gitlab for hosting these tutorials and associated code has simplified the process of sharing material.
To date, however, there has been little coordination in this effort.
The coverage of topics in the available material is not uniform.
Moreover, the underlying principles and foundations of statistics are often not covered in favor of the introduction of commonly used software tools and algorithms.
For the case of algorithmic design and optimization, there has been substantially less progress in training the community.
Instead, the primary focus being the development of introductory materials such as the Software and Data carpentry \citep{data_carpentry}. 
We have started to make progress in providing an educational foundation in statistics and algorithms, but it is not uniformly available across our community --- with significantly less access at smaller colleges and in underrepresented communities.
We, therefore, recommend the development and support of a common/open set of educational resources that can be used in teaching statistics, and algorithms, and machine or computational learning. Determining what constitutes an appropriate curriculum will be a balance between providing the foundations of statistics and algorithmic design appropriate for the broader science community and teaching specialized skills (e.g. optimization, compilers) that may benefit a smaller, but crucial, set of researchers who will engage in the development and implementation of computing and software frameworks.

\section{Ethical Implications for Contributing to the Advancement of Algorithms and Statistical Models}

A new era in algorithmic development and application in social contexts has arrived. 
This era is driven primarily and technically by artificial intelligence, increasingly accessible high-performance computing hardware, and large data sets. 
AI algorithms are a new tool and field of study with significantly cross-cutting capabilities: they are and will likely continue to have a great influence both on scientific discovery and systems of social and cultural relevance. 

AI has begun to play significant roles in the areas of news and media \citep[e.g., deep fakes;][]{Villasenor2019}, workforce development \citep[e.g., automation across multiple sectors;][]{Manyika2018}, criminal justice \citep[e.g.,][]{Angwin2016}, and privacy \citep[e.g., surveillance, facial recognition;][]{MacCarthy2019}.
Given the broad applicability of these algorithms and the rapid incorporation of them into many computational systems, when generally useful advancements are likely to be adopted readily.
Therefore, those who play a role in the advancement of these algorithms may be contributing to changes outside of a narrow technical purpose. 
For example, astronomers who seek to use new algorithms and computing paradigms are poised to contribute to their development so that the new tools are useful in our field of study. 
As we make these contributions, we are faced with ethical questions regarding the potentially broad impact of this work. 

This is not the first time that space scientists, astronomers, and physicists have played roles in developing new technologies or facilitating their application in contexts outside the field proper (e.g., the global positioning satellite system, nuclear technology).
As users, facilitators, and developers of these technologies, astronomers should do more in the petabyte era to reckon with these activities and their broad societal impacts.

\newpage
\bibliographystyle{yahapj}
\bibliography{main}

\begin{thebibliography}{}
\providecommand\natexlab[1]{#1}
\providecommand\JournalTitle[1]{#1}

\bibitem[{{Alsing} {et~al.}(2018){Alsing}, {Wandelt}, \&
  {Feeney}}]{2018MNRAS.477.2874A}
{Alsing}, J., {Wandelt}, B., \& {Feeney}, S. 2018,
  \href{http://dx.doi.org/10.1093/mnras/sty819}{\JournalTitle{\mnras}, 477,
  2874}

\bibitem[{Angwin {et~al.}(2018)Angwin, Larson, Mattu, \& Kirchner}]{Angwin2016}
Angwin, J., Larson, J., Mattu, S., \& Kirchner, L. 2018,
  \href{https://www.propublica.org/article/machine-bias-risk-assessments-in-criminal-sentencing}{\JournalTitle{ProPublica}}

\bibitem[{{Astropy Collaboration}(2019)}]{astropy_tutorials}
{Astropy Collaboration}. 2019

\bibitem[{{Baron} \& {Poznanski}(2017)}]{2017MNRAS.465.4530B}
{Baron}, D., \& {Poznanski}, D. 2017,
  \href{http://dx.doi.org/10.1093/mnras/stw3021}{\JournalTitle{\mnras}, 465,
  4530}

\bibitem[{{Bauer} {et~al.}(2019){Bauer}, {Bellm}, {Bolton}, {Chaudhuri},
  {Connolly}, {Cruz}, {Desai}, {Drlica-Wagner}, {Economou}, \&
  {Gaffney}}]{2019arXiv190505116B}
{Bauer}, A.~E., {Bellm}, E.~C., {Bolton}, A.~S., {et~al.} 2019,
  \JournalTitle{arXiv e-prints}, arXiv:1905.05116

\bibitem[{{Ivezi{\'c}} {et~al.}(2014){Ivezi{\'c}}, {Connolly}, {Vanderplas}, \&
  {Gray}}]{astroMLText}
{Ivezi{\'c}}, {\v Z}., {Connolly}, A., {Vanderplas}, J., \& {Gray}, A. 2014,
  Statistics, Data Mining and Machine Learning in Astronomy (Princeton
  University Press)

\bibitem[{Jao {et~al.}(2018)Jao, Henry, Gies, \& Hambly}]{Jao_2018}
Jao, W.-C., Henry, T.~J., Gies, D.~R., \& Hambly, N.~C. 2018,
  \href{http://dx.doi.org/10.3847/2041-8213/aacdf6}{\JournalTitle{The
  Astrophysical Journal}, 861, L11}

\bibitem[{Kluyver {et~al.}(2016)Kluyver, Ragan-Kelley, P{\'e}rez, Granger,
  Bussonnier, Frederic, Kelley, Hamrick, Grout, Corlay, Ivanov, Avila, Abdalla,
  \& Willing}]{jupyter}
Kluyver, T., Ragan-Kelley, B., P{\'e}rez, F., {et~al.} 2016, in Positioning and
  Power in Academic Publishing: Players, Agents and Agendas, ed. F.~Loizides \&
  B.~Schmidt, IOS Press, 87

\bibitem[{Kohl(2015)}]{R:Kohl:2015en}
Kohl, M. 2015, Introduction to statistical data analysis with {R} (London:
  bookboon.com)

\bibitem[{MacCarthy(2019)}]{MacCarthy2019}
MacCarthy, M. 2019,
  \href{https://www.brookings.edu/blog/techtank/2019/04/01/how-to-address-new-privacy-issues-raised-by-artificial-intelligence-and-machine-learning/}{\JournalTitle{Brookings
  Institute}}

\bibitem[{Manyika \& Sneader(2018)}]{Manyika2018}
Manyika, J., \& Sneader, K. 2018,
  \href{https://www.mckinsey.com/featured-insights/future-of-work/ai-automation-and-the-future-of-work-ten-things-to-solve-for}{\JournalTitle{McKinsey
  Global Institute}}

\bibitem[{{Suzuki} \& {Fukugita}(2018)}]{2018AJ....156..219S}
{Suzuki}, N., \& {Fukugita}, M. 2018,
  \href{http://dx.doi.org/10.3847/1538-3881/aac88b}{\JournalTitle{\aj}, 156,
  219}

\bibitem[{Villasenor(2019)}]{Villasenor2019}
Villasenor, J. 2019,
  \href{https://www.brookings.edu/blog/techtank/2019/02/14/artificial-intelligence-deepfakes-and-the-uncertain-future-of-truth/}{\JournalTitle{Brookings
  Institute}}

\bibitem[{{von Hardenberg} {et~al.}(2019){von Hardenberg}, Obeng, Pawlik,
  Pletzer, Shiklomanov, Fouilloux, Wright, Fournier, Marwick, Brown, Johnson,
  Voter, Hulshof, Bahlai, Shaw, Li, Bouquin, Stubbs, Quinn, Vanichkina,
  Fishman, Wilson, Hart, Hannon, Sügis, Strauss, Gan, Becker, White,
  Rodriguez-Sanchez, Michonneau, Boehm, {GMoncrieff}, Ye, Dashnow, Lapp,
  {JSurman}, Ashander, Byrnes, Hollister, Chen, Dunic, {Jon}, Keane, Stachelek,
  Herr, Mislan, Woo, Cranston, Jordan, Ram, Hertweck, Todd-Brown, Lotterhos,
  Peck, Direk, Hall, Tylén, Chatzidimitriou, Deer, Gatto, Wasser, Tarkowski,
  Breckels, Foos, Chiapello, Robinson, Akenbrand, Kuzak, Grenié, Grenié,
  Salmon, Duffy, Koontz, Johnston, Marino, Carchedi, Burge, Lijnzaad, Lijnzaad,
  Peek, Supp, Taylor, Labou, Pederson, Webster, Reiter, Sandmann, Teal,
  Furnass, Pearse, Li, Lapp, {ab604}, {ashander}, {cengel}, Seok, {sfn_brt}, \&
  {suparee}}]{data_carpentry}
{von Hardenberg}, A., Obeng, A., Pawlik, A., {et~al.} 2019, Data Carpentry: R
  for data analysis and visualization of Ecological Data

\end{thebibliography}

\end{document}